\documentclass[conference]{IEEEtran}
\IEEEoverridecommandlockouts

\usepackage{cite}
\usepackage{amsmath,amssymb,amsfonts}
\usepackage{algorithmic}
\usepackage{graphicx}
\usepackage{textcomp}
\usepackage{tabularx}
\usepackage{multirow}
\usepackage{makecell}
\usepackage{orcidlink}
\usepackage{booktabs}
\usepackage{array}

\usepackage{tcolorbox} 

\def\BibTeX{{\rm B\kern-.05em{\sc i\kern-.025em b}\kern-.08em
    T\kern-.1667em\lower.7ex\hbox{E}\kern-.125emX}}
\begin{document}

\title{Large Language Model-Based Framework for Explainable Cyberattack Detection in Automatic Generation Control Systems}

\author{
Muhammad Sharshar\orcidlink{0009-0003-8396-842X}, 
Ahmad Mohammad Saber\orcidlink{0000-0003-3115-2384}, \IEEEmembership{Member, IEEE},  
Davor Svetinovic\orcidlink{0000-0002-3020-9556}, \IEEEmembership{Senior Member, IEEE},\\
Amr M. Youssef\orcidlink{0000-0002-4284-8646}, \IEEEmembership{Senior Member, IEEE}, 
Deepa~Kundur\orcidlink{0000-0001-5999-1847},~\IEEEmembership{Fellow,~IEEE},
and Ehab F. El-Saadany\orcidlink{0000-0003-0172-0686}, \IEEEmembership{Fellow, IEEE}
\thanks{Muhammad Sharshar and Davor Svetinovic are with the Department of Computer Science, College of Computing and Mathematical Sciences, Khalifa University, Abu Dhabi, UAE (emails: \href{mailto:mohamed.sharshar@ku.ac.ae}{mohamed.sharshar@ku.ac.ae}, \href{mailto:davor.svetinovic@ku.ac.ae}{davor.svetinovic@ku.ac.ae}).}
\thanks{Ahmad Mohammad Saber and Deepa Kundur are with the ECE Department, University of Toronto, Toronto, ON, Canada (emails: \href{mailto:ahmad.m.saber@ieee.org}{ahmad.m.saber@ieee.org}, \href{mailto:dkundur@ece.utoronto.ca}{dkundur@ece.utoronto.ca}).}
\thanks{Amr Youssef is with Concordia Institute for Information Systems Engineering (CIISE), Concordia University, Montreal, QC, Canada (email: \href{mailto:youssef@ciise.concordia.ca}{youssef@ciise.concordia.ca}).}
\thanks{Ehab El-Saadany is with the Department of Electrical Engineering, College of Engineering and Physical Sciences, Khalifa University, Abu Dhabi, UAE (email: \href{mailto:ehab.elsadaany@ku.ac.ae}{ehab.elsadaany@ku.ac.ae}).}
}

\maketitle

\begin{abstract}

The increasing digitization of smart grids has improved operational efficiency but also introduced new cybersecurity vulnerabilities, such as  False Data Injection Attacks (FDIAs) targeting Automatic Generation Control (AGC) systems. While machine learning (ML) and deep learning (DL) models have shown promise in detecting such attacks, their opaque decision-making limits operator trust and real-world applicability. This paper proposes a hybrid framework that integrates lightweight ML-based attack detection with natural language explanations generated by Large Language Models (LLMs). Classifiers such as LightGBM achieve up to 95.13\% attack detection accuracy with only 0.004 s inference latency. Upon detecting a cyberattack, the system invokes LLMs—including GPT-3.5 Turbo, GPT-4 Turbo, and GPT-4o mini—to generate human-readable explanation of the event. Evaluated on 100 test samples, GPT-4o mini with 20-shot prompting achieved 93\% accuracy in identifying the attack target, a mean absolute error of 0.075 pu in estimating attack magnitude, and 2.19 seconds mean absolute error (MAE) in estimating attack onset. These results demonstrate that the proposed framework effectively balances real-time detection with interpretable, high-fidelity explanations, addressing a critical need for actionable AI in smart grid cybersecurity.

\end{abstract}


\section{Introduction}

Smart grids represent a transformative evolution of traditional power systems, integrating advanced sensing, communication, and control technologies to enhance efficiency and reliability \cite{manias2024trends}. Automatic Generation Control (AGC) systems play a critical role in maintaining frequency stability and balancing power across interconnected areas  \cite{kundur1994power}. However, the increased digitization and connectivity of these systems expose them to significant cybersecurity threats, including {False Data Injection Attacks (FDIAs)} that manipulate targeted component's measurements to disrupt grid operation~\cite{khalaf2018joint}. As a response, {machine learning (ML)} and {deep learning (DL)} techniques have gained traction for cyberattack detection, offering high accuracy in identifying subtle cyber-physical attacks~\cite{ozay2016ml}.

Yet, a fundamental challenge persists: the lack of transparency in ML/DL models, particularly neural networks, makes it difficult for power system operators to trust and act on their outputs. 
Experts may not fully trust the outputs of machine learning-based algorithms, particularly in scenarios where operational reliability is critical, a frequent requirement in the energy sector. 
In high-stakes environments such as AGC, where incorrect or unexplained decisions may trigger operational instability, \textit{explainability is essential} for real-world adoption.

While existing explainability techniques like SHAP and LIME \cite{xu2019explainable} offer  insights into ML predictions, they often fall short of producing operator-friendly interpretations. {Large Language Models (LLMs)}, in contrast, offer a unique opportunity to bridge this gap. With their ability to process domain-specific information, infer contextual meaning, and generate {natural language justifications}, LLMs can support {human-understandable explanations} of ML-based decisions, an approach increasingly explored in cybersecurity and industrial AI~\cite{lundberg2017shap,doshi2021explainable}.

Large Language Models (LLMs) are increasingly explored in power systems for their ability to generate interpretable insights from complex data, yet their application in smart grid cybersecurity remains limited \cite{shi2024review}. Chen et al. \cite{Chen2024} developed an LLM-based framework for emergency control, using ChatGPT and fine-tuned LLMs to interpret alarms and recommend actions with 92\% accuracy. However, it focuses on control rather than cybersecurity. Liu et al. \cite{Liu2024} employed LLMs like LLaMA for power system analysis, reducing report generation time by 15\%, but did not address real-time threats. Zhang et al. \cite{Zhang2023} used GPT-3.5 for anomaly detection in distribution networks, improving operator comprehension by 20\% through textual summaries.
While prior works demonstrate the potential of LLMs in power system applications, the integration of LLMs with real-time ML classifiers for explainable FDIA detection, as proposed in this paper, remains underexplored. %

This paper introduces a hybrid framework that combines fast detection of FDIAs using lightweight ML models with  explanations generated by LLMs. A key feature of the proposed framework is its alignment with power system operator workflows. 
Once a cyberattack is detected, the system invokes an LLM to interpret the ML model's decision  and associated signal features to produce structured, human-readable reports. These reports include estimated attack start time, affected variables, and statistical justifications, presented in operator-friendly narratives. 
Such reports enable operators to interpret alarms with confidence, supporting decision-making regarding mitigation, incident logging, or further investigation. 
By aligning with the operational needs of power system personnel and reducing the interpretability gap, the proposed framework improves decision-making, enhances operator trust and promotes actionable use of AI tools in critical power infrastructure.

The main contributions of this work can be summarized as:
\begin{itemize}
    \item A novel integration of powerful ML classifiers, such as LightGBM and XGBoost, and LLMs for detecting and explaining attacks, respectively, against AGC systems.
    \item A comprehensive evaluation of few-shot learning configurations for LLMs in smart grid cybersecurity.
\end{itemize}

\begin{figure}[t!]
\centering
\includegraphics[width=1.0\linewidth,keepaspectratio]{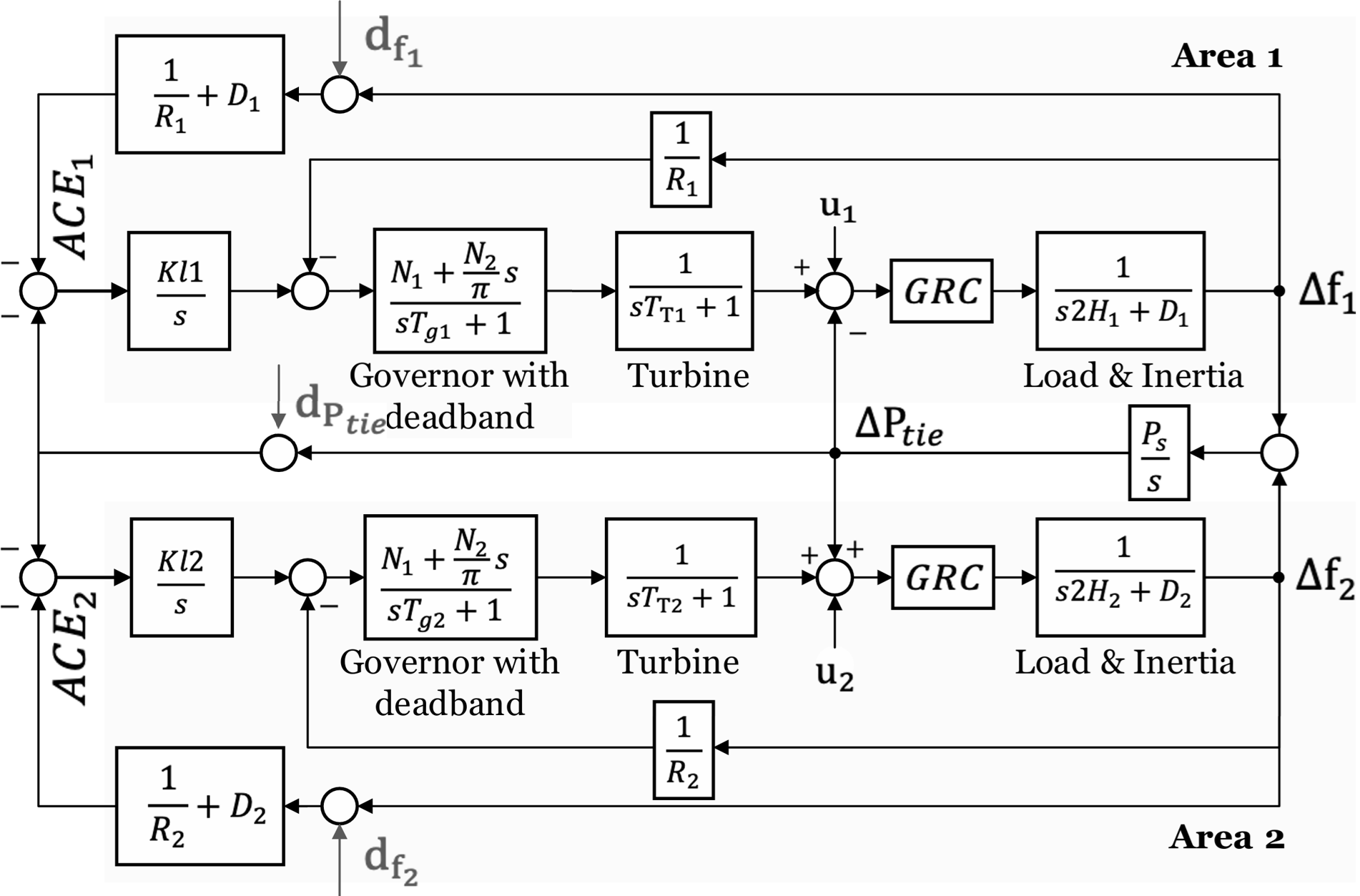}
\caption{Block diagram of test system showing non-linearities.}\label{fig:2area}
\end{figure}

\section{Case Study: Non-Linear Two-Area AGC System }
To investigate the detection and explanation of cyber-attacks in Automatic Generation Control (AGC) systems,  a two-area AGC model is utilized to generate synthetic time-series signals.  The two-area system, a standard framework for studying interconnected power systems \cite{kundur1994power} whose block diagram is shown in Fig. \ref{fig:2area}, comprises two control areas, each with a governor, turbine, and load, connected via a tie-line. The dynamics of the system are modeled using differential equations that describe frequency deviations and tie-line power flows \cite{elgerd1982electric}.

For each area \( i \), i = 1, 2, the frequency deviation \( \Delta f_i \) is governed by the swing equation:
\begin{equation}
\frac{2H_i}{\omega_0} \frac{d \Delta f_i}{dt} = \Delta P_{m,i} - \Delta P_{L,i} - D_i \Delta f_i - \Delta P_{\text{tie}},
\label{eq:swing}
\end{equation}
where \( H_i \) is the inertia constant, \( \omega_0 \) is the nominal frequency, \( \Delta P_{m,i} \) is the mechanical power change, \( \Delta P_{L,i} \) is the load change, \( D_i \) is the load damping coefficient, and \( \Delta P_{\text{tie}} \) is the tie-line power flow deviation. The tie-line power flow between areas is given by:
\begin{equation}
\Delta P_{\text{tie}} = \frac{T_{12}}{s} (\Delta f_1 - \Delta f_2),
\label{eq:tie}
\end{equation}
where \( T_{12} \) is the synchronizing coefficient, and \( s \) represents the Laplace operator. The AGC system adjusts the mechanical power \( \Delta P_{m,i} \) based on the Area Control Error (ACE), defined as:
\begin{equation}
\text{ACE}_i = B_i \Delta f_i + \Delta P_{\text{tie}},
\label{eq:ace}
\end{equation}
where \( B_i \) is the frequency bias factor \cite{kundur1994power}.

\section{Methodology and Experimental Setup}

\begin{figure*}[htbp]
    \centering
    \includegraphics[width=1\textwidth]{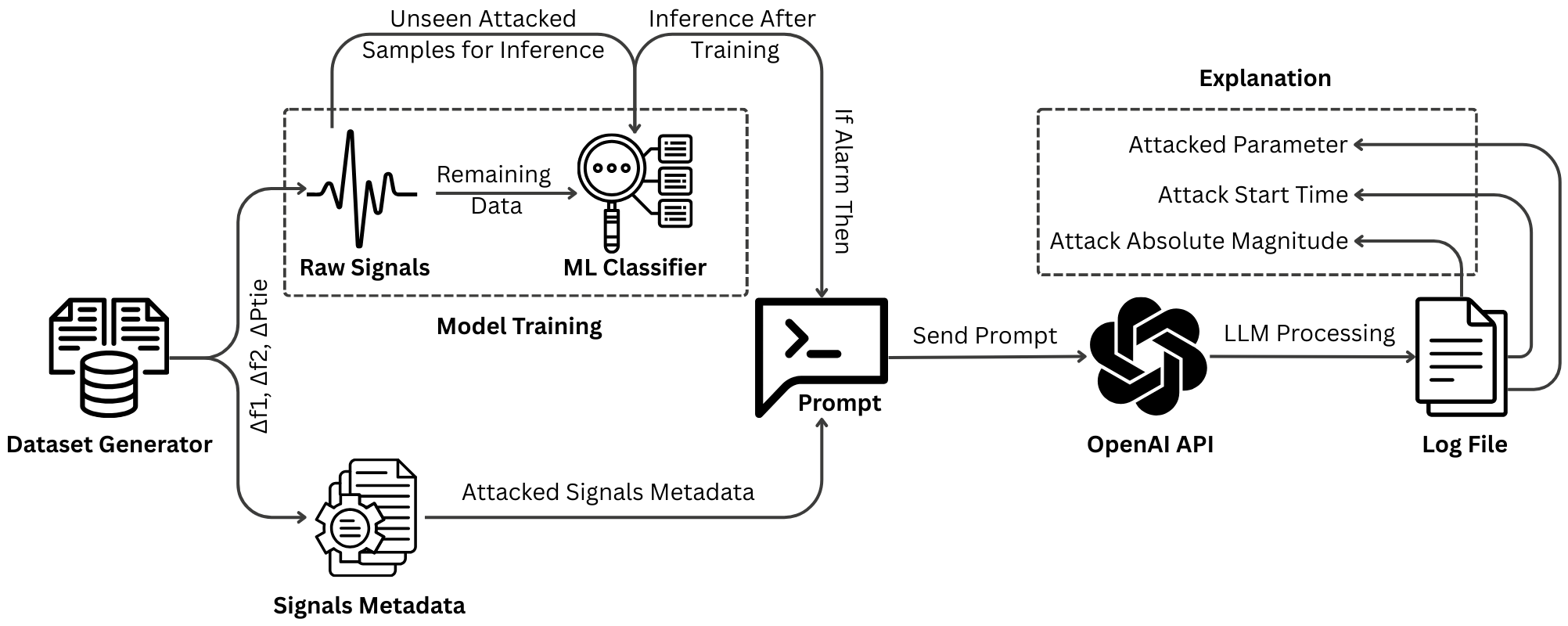}
    \caption{Overview of the proposed ML–LLM framework for cyberattack explanation. The ML classifier labels incoming AGC signals, and the LLM generates human-readable explanation based on the classifier outputs and signal metadata.}
    \label{fig:system}
\end{figure*}

This study presents FDIA detection–explanation framework, where attack detection is designed to be fast and accurate, followed by  explanation and justification for forensic logging and future inspection. The proposed framework, illustrated in Figure~\ref{fig:system}, consists of four phases: dataset generation, ML classifier training, prompt construction, and LLM-based explanation.

\subsection{Dataset Generation}

\begin{table}[!t]
\caption{Key System Parameters\label{tab:system_params}}
\centering
\footnotesize
\renewcommand{\arraystretch}{1.3} 
\begin{tabular}{c|c|l}
\toprule
\textbf{Parameter} & \textbf{Value(s)} & \textbf{Description} \\ \hline
$H_1, H_2$          & 5, 4        & Inertia constants (s)               \\
$D_1, D_2$          & 0.6, 0.3    & Damping coefficients (pu/Hz)        \\
$B_1, B_2$          & 20.6, 16.3  & Frequency bias (pu/Hz)              \\
$T_{g,1}, T_{g,2}$  & 0.2, 0.3    & Governor time constants (s)         \\
$T_{t,1}, T_{t,2}$  & 0.5, 0.6    & Turbine time constants (s)          \\
$R_1, R_2$          & 0.05, 0.0625 & Speed regulation (pu)              \\
$T_{12}$            & 2           & Synchronizing coefficient (pu)      \\
$\text{GDB}$        & 0.06        & Governor deadband (\%)             \\
$GRC_1, GRC_2$      & $\pm3$      & Gen.\ rate constraints (pu/min)     \\
$K_{\text{i},1},K_{\text{i},2}$ & 0.3, 0.3 & AGC integral gains (-)  \\
\bottomrule
\end{tabular}
\end{table}

\begin{figure}[htbp]
    \centering
    \includegraphics[width=\linewidth]{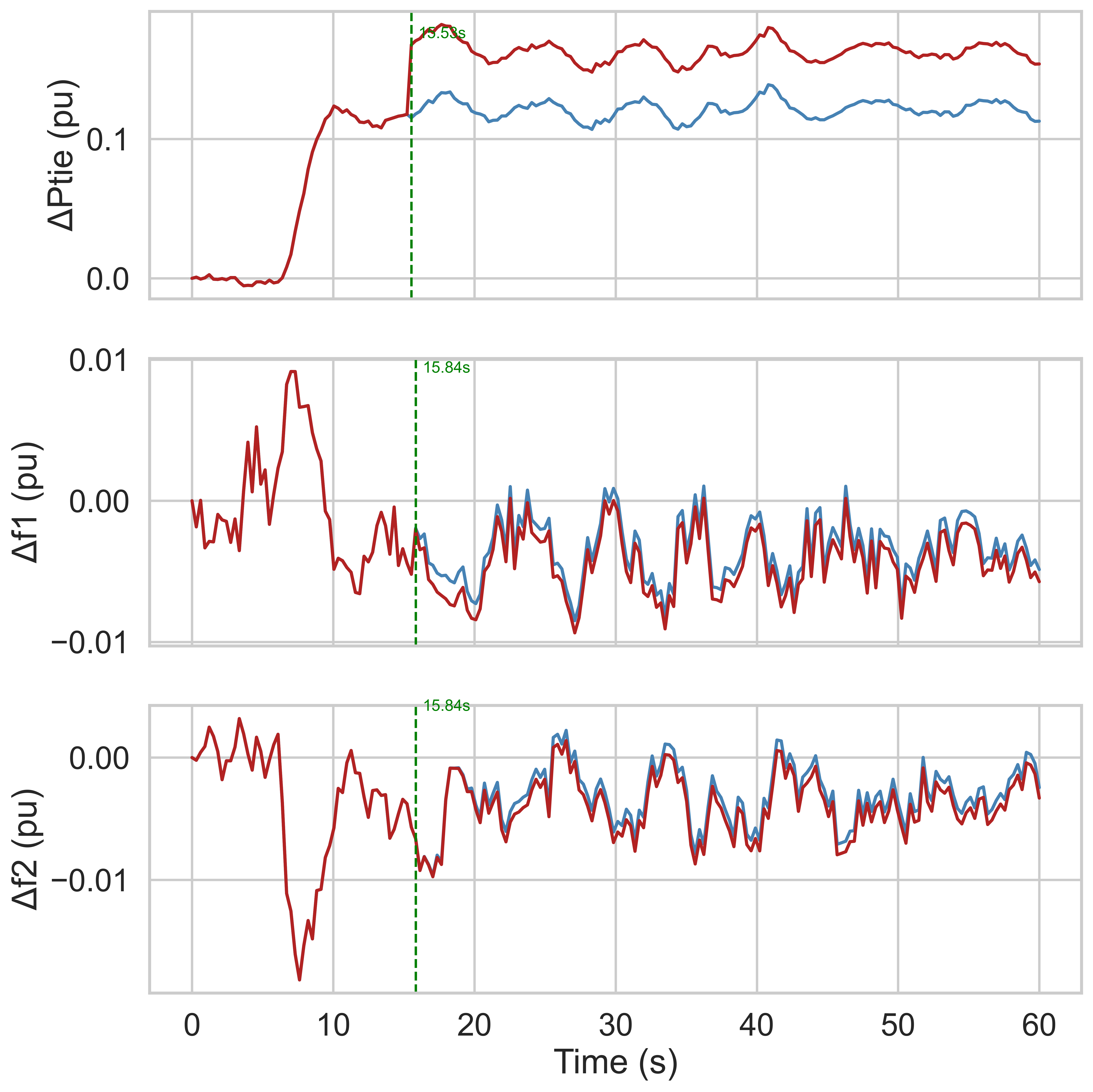}
    \caption{Dataset sample showing raw signals of \( \Delta f_1 \), \( \Delta f_2 \), and \( \Delta P_{\text{tie}} \), with the normal variant in blue and the variant under FDIA shown in red.}
    \label{fig:dataSample}
\end{figure}

The two-area power system with AGC \cite{kundur1994power}, whose block diagram is illustrated in Fig. \ref{fig:2area}  is simulated.
The system is simulated including the AGC nonlinearities.
Table~\ref{tab:system_params} summarizes the simulation parameters and constants. Each sample consists of multivariate time-series data—specifically three signals ($\Delta f_1$, $\Delta f_2$, and $\Delta P_{\text{tie}}$)—accompanied by metadata. This metadata includes statistical features (mean, standard deviation, skewness) as well as simulation parameters such as disturbance time, noise levels, and attack characteristics for the FDIA samples.

Each sample spans 60 seconds, sampled at 0.3-second intervals, yielding 200 data points per sample, adequate for capturing system dynamics without requiring AGC oversampling. Disturbance magnitudes were drawn from a zero-mean normal distribution with a standard deviation of 0.02 p.u. Attacks were introduced post-disturbance to simulate realistic interference scenarios. Based on empirical evaluation, attack parameters \(f_f\) and \(f_i\) were sampled from a normal distribution centered at \(-0.11\) with a standard deviation of 0.02, as these values were found to yield the most effective attacks. To emulate real-world uncertainties, white Gaussian noise with zero mean and standard deviation \(10^{-6}\) was added to model both process and measurement noise, representing stochastic system behavior and sensor inaccuracies, respectively. An ACE limit was enforced to constrain post-attack deviations: 0.5 for subtle attacks and 1.0 for more noticeable ones. Attacks were rescaled accordingly if they exceeded these thresholds, ensuring a wide range of difficulty levels in the generated data.

This methodology resulted in a balanced dataset of 10{,}000 samples, equally divided between normal and attack cases, with varied configurations. Unlike previous studies that used fixed timing, both disturbance and attack initiation times were randomized within the first 30 seconds to avoid static patterns. The simulation considered both linearized and nonlinear system behaviors, the latter incorporating generation rate constraints (GRC), deadband effects, and time delays. Figure~\ref{fig:dataSample} presents a sample visualization, illustrating how attack-induced deviations can be subtle, with an attack magnitude of 0.2138pu, an attack start time of 15 seconds, and $\Delta P_{\text{tie}}$ as the targeted parameter. This synthetic simulation approach was necessary due to the unavailability of real-world measurement data for security reasons. Even if such data were accessible, they would likely lack the variability needed to comprehensively evaluate FDIA detection methods, a limitation also acknowledged in prior studies that similarly relied on synthetic datasets.

\subsection{Machine Learning Models Training}

Using LLMs directly for signal classification presents challenges in both local and online deployment scenarios. LLMs demand computational resources proportional to the number of input tokens. 
To overcome these limitations, classification is performed by lightweight local ML models, while the LLM is employed  to generate explanations based on summarized model outputs.
In other words, in this work, the LLM is used as an explanation generator rather than a classifier.
Following this approach, multiple ML models are evaluated, and the best-performing one is selected for the next phase. XGBoost and LightGBM are employed due to their well-established performance on structured time-series and fast inference capabilities. 
Random Forests (RF) is also considered for its robustness and effectiveness in similar tasks.
A subset of 200 samples from each of the normal and attacked datasets is held out exclusively for LLM evaluation. From the remaining data, 70\% is used for training and validation, while 30\% is reserved for testing. These models are trained on signal triplets and are tasked with binary classification, labeling each sample as either normal or under FDIA.

\subsection{LLM API Setup \& Prompt Structure}

For the explanation generation task, the OpenAI ChatGPT API was employed. The \texttt{gpt-3.5-turbo}, \texttt{gpt-4-turbo}, and \texttt{gpt-4o mini} models were selected based on their strengths in reasoning, accessibility, and performance-cost trade-offs. \texttt{gpt-3.5-turbo} was included as a strong baseline due to its wide availability and low inference cost, making it suitable for large-scale or cost-sensitive deployments. \texttt{gpt-4-turbo} was evaluated for its enhanced reasoning capabilities, particularly in tasks requiring logical deduction, pattern recognition, and structured output. \texttt{gpt-4o mini} was chosen for its optimized balance of accuracy, latency, and computational efficiency. Together, these models provide a representative spectrum for evaluating LLM behavior under varying performance and resource constraints. To ensure reproducibility across runs and users, all API calls were configured with a fixed seed value, enabling deterministic behavior when supported. Furthermore, the temperature parameter was set to 0.0 to minimize randomness in the generated outputs. All other generation parameters were retained at their default values unless otherwise specified.

Each API request consists of two components: a {system prompt} and a {query}. The system prompt provides global task instructions and defines the role of the LLM as a cybersecurity analyst responsible for explaining ML model predictions related to AGC signal anomalies. This role-based specification helps constrain the model's behavior toward the intended analytical function. The prompt includes background information and detailed system parameter specifications for the two-area AGC system to ensure the model has the necessary context to reason about control dynamics. It also outlines the nature of the input signals—$\Delta f_1$, $\Delta f_2$, and $\Delta P_{\text{tie}}$—and describes the accompanying metadata features, such as statistical summaries (mean, standard deviation, skewness, slope, min, max) and noise settings. The required output format is explicitly defined as structured JSON with natural language justifications to ensure consistency and interpretability. A configurable number of few-shot examples can be included to guide the model’s reasoning through representative cases.

The {query} section contains the metadata for an unseen FDIA test sample to be inferred
for the three signals. 
In addition, the query is appended with the inference results from the best-performing ML classifier, specifically including the predicted label, confidence score, and class probabilities. The query is deliberately structured to exclude ground-truth attack characteristics (e.g., start time, magnitude, and target), allowing the LLM to independently infer these hidden attributes. Notably, only samples classified as attacks are considered for explanation, as the objective is to interpret alarms; applying the same process to normal samples would introduce unnecessary overhead. These inferences are then compared against the true values for evaluation of explanation quality. Finally, care is taken to ensure that the complete prompt—including the system prompt, few-shot examples, and query—remains within the token limitations of the selected LLM model, as exceeding these limits may result in input truncation or degraded response fidelity.

\begin{figure}[t]
    \centering
    \includegraphics[width=1\linewidth]{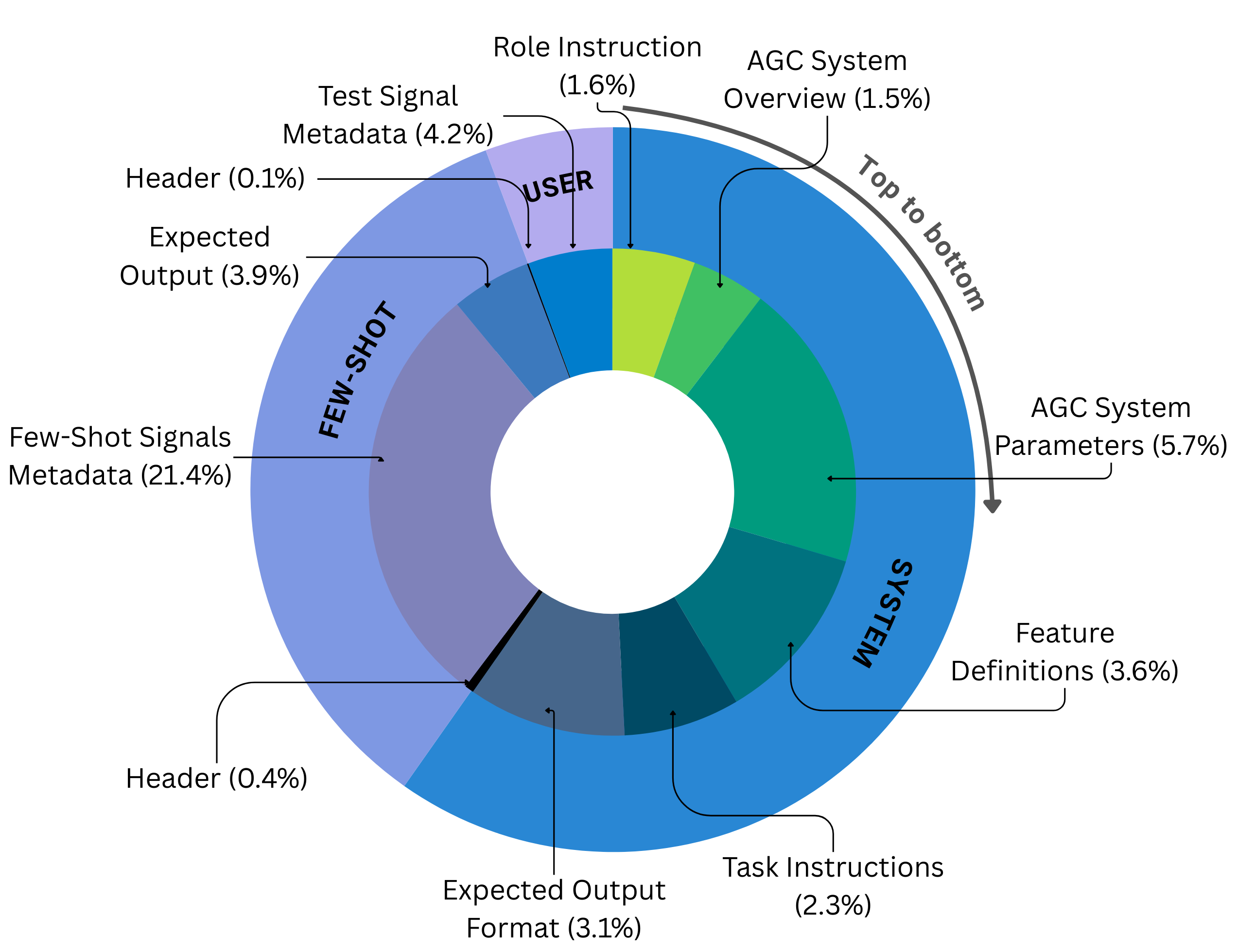}
    \caption{Token composition of a complete 5-shot prompt provided to the LLM}
    \label{fig:token_breakdown}
\end{figure}

To better illustrate the structure and relative weight of each component in the constructed prompt, Fig.~\ref{fig:token_breakdown} presents the token distribution. However, the inclusion of few-shot samples introduces a noticeable overhead that increases linearly with the number of shots. Since the API is stateless, these examples must be provided with every call, as each request is processed independently. Despite this additional cost, few-shot prompting, in this case, remains more practical than fine-tuning a model, which would require specialized hardware and significantly higher computational resources for deployment.

\section{Results \& Discussion}
To validate the proposed explainable FDIA detection pipeline, the classification performance of selected ML models is reported, along with the interpretive capabilities of different LLM configurations. The results are analyzed to highlight both the accuracy of detection and the practical utility of the generated explanations for forensic analysis.

\begin{table*}[t!]
\centering
\caption{Performance of ML Models in Attack Detection}
\label{tab:ml-performance}
\renewcommand{\arraystretch}{1.3} 
\begin{tabular}{lccccc}
\toprule
\textbf{Model} & \textbf{Accuracy} & \textbf{Recall} & \textbf{Precision} & \textbf{F1 Score} & \textbf{Latency (s)} \\
\midrule
LightGBM       & \textbf{0.9513}   & 0.9160          & 0.9857             & 0.9496             & \textbf{0.004} \\
XGBoost        & 0.9467            & 0.9187          & 0.9732             & 0.9451             & 0.007 \\
Random Forest  & 0.9303            & 0.8647          & 0.9954             & 0.9254             & 0.013 \\
\bottomrule
\end{tabular}
\end{table*}

\begin{table*}[htbp]
\caption{Performance of LLM Models in Attack Explanation}
\centering
\renewcommand{\arraystretch}{1.1}
\setlength{\tabcolsep}{5pt}
\begin{tabular}{
>{\centering\arraybackslash}p{2.4cm} 
>{\centering\arraybackslash}p{1.3cm} 
>{\centering\arraybackslash}p{2.2cm} 
>{\centering\arraybackslash}p{2.2cm} 
>{\centering\arraybackslash}p{2.2cm} 
>{\centering\arraybackslash}p{2.2cm} 
}
\toprule
\textbf{Model} & \textbf{Shots} & \textbf{Accuracy of Attack Target(\%)} & \textbf{MAE of Attack Magnitude} & \textbf{MAE of Attack Time} & \textbf{Latency (s)} \\
\midrule
\multirow{4}{*}{GPT-3.5 Turbo}
& 0  & 45.00 & 0.08500 & 3.83 & 1.730 \\
& 5  & 88.00 & 0.02648 & 2.37 & 2.224 \\
& 10 & 73.00 & 0.01952 & 2.42 & 2.670 \\
& 20 & 65.00 & 0.01989 & 2.05 & 2.980 \\
\midrule
\multirow{4}{*}{GPT-4 Turbo}
& 0  & 41.00 & 0.12331 & 2.92 & 5.099 \\
& 5  & 82.00 & 0.07280 & 2.61 & 5.368 \\
& 10 & 87.00 & 0.04338 & 2.35 & 5.141 \\
& 20 & 87.00 & 0.03971 & 2.12 & 6.271 \\
\midrule
\multirow{4}{*}{GPT-4o mini}
& 0  & 83.00 & 0.14276 & 3.87 & 5.560 \\
& 5  & 90.00 & 0.12106 & 2.48 & 5.740 \\
& 10 & 89.00 & 0.08979 & 2.21 & 5.990 \\
& 20 & 93.00 & 0.07519 & 2.19 & 6.870 \\
\bottomrule
\end{tabular}
\label{tab:llm_performance_baseline}
\end{table*}

\subsection{Machine Learning Models Classification Performance}

Table~\ref{tab:ml-performance} illustrates the performance of the models, which were trained and evaluated using standard classification metrics—accuracy, precision, recall, and F1-score \cite{saber2024unmasking}—on a test set representing both normal and FDIA conditions. This also shows that the generated data is challenging because of the randomized attack times and other factors. The objective was not only to achieve high classification performance but also to generate informative outputs (e.g., confidence scores) to guide the subsequent LLM explanation phase. While most models exhibited strong performance, LightGBM outperformed the others with higher detection accuracy and higher precision leading to fewer false alarms, and was therefore selected for use in prompt construction. Hyperparameter tuning was performed for all models using Optuna \cite{akiba2019optuna} to ensure near-optimal results.

\subsection{Explanation Quality Assessment}
Table~\ref{tab:llm_performance_baseline}
compares the average performance of multiple test runs on 100 FDIA samples, selected to enable manual inspection, across three LLM variants: GPT-3.5 Turbo, GPT-4 Turbo, and GPT-4o mini. The comparison spans varying shot settings (0, 5, 10, and 20), highlighting both baseline results and the corresponding changes.
The table provides insights into how each model responds to the integration of learned signal features with respect to classification accuracy, regression error, and latency. 
We observe inconsistency in the improvement of GPT-3.5's accuracy as the number of shots increases.  
In contrast, GPT-4 Turbo exhibits relatively stable performance across all shots, with minor degradation in classification metrics and manageable variations in (Mean Absolute Error) MAE. This resilience highlights GPT-4’s robustness to additional features, likely due to its stronger architectural capacity and alignment. GPT-4o mini, however, demonstrates a more nuanced pattern with minor drops in accuracy at some shot levels
Furthermore, GPT-3.5 unexpectedly exhibits smaller MAEs for both attack magnitude and time, at 10 shots, with all models exhibiting attack magnitude MAE of less than 0.02, and  attack time of less than 2.2.
Finally, we observe that the latency increases with the LLM's complexity. However, the latency of all models is acceptable compared to the time scale of AGC systems, as modern AGC system collect measurement every few minutes. Recall also that attack detection is not impacted by this latency as the attack detection is performed by the ML model whose inference time is in the range of milliseconds as depicted in Table \ref{tab:ml-performance}.

\begin{figure}[htbp]
    \centering
    \includegraphics[width=\linewidth]{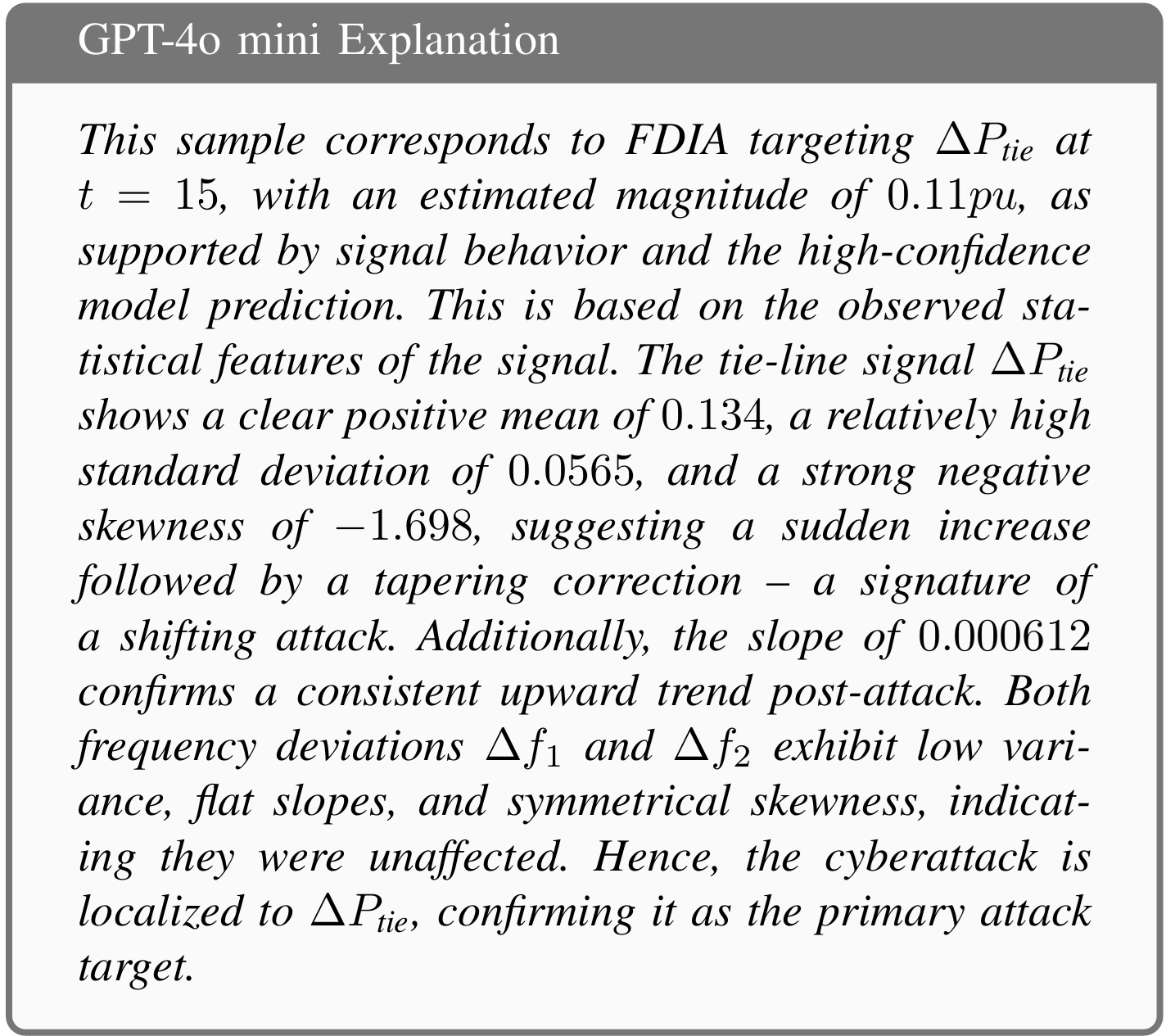}
    \caption{GPT-4o mini explanation for the sample in Figure~\ref{fig:dataSample}.}
    \label{fig:gpt4oExplanation}
\end{figure}

Evaluating the sample shown in Figure~\ref{fig:dataSample}, GPT-4o mini with 20-shot prompting provided  explanation shown in Fig. \ref{fig:gpt4oExplanation}.
This response demonstrates the model's ability to accurately interpret signal dynamics and justify its classification with detailed statistical and human readable reasoning.

\section{Conclusion and Future Work}
This paper presented an ML-LLM framework that integrates lightweight ML classifiers with LLMs to detect and explain FDIAs in AGC systems. By deploying ML models locally for real-time classification and leveraging LLMs for  explanation, the approach ensures low-latency detection, transparency, and data privacy. Experimental results demonstrate high detection accuracy and effective explanation generation across various LLM configurations and prompt designs. 

Future work will explore lightweight, locally deployable LLMs for on-premise inference, adaptive retraining strategies for ML classifiers, and decentralized architectures that support explainability without compromising real-time performance. It will be also interesting to
optimize prompt and
address potential LLM drawbacks (e.g., hallucinations, biases), which could generate misleading explanations in such critical applications.

\bibliographystyle{IEEEtran} 

\bibliography{References}

\begin{thebibliography}{10}
\providecommand{\url}[1]{#1}
\csname url@samestyle\endcsname
\providecommand{\newblock}{\relax}
\providecommand{\bibinfo}[2]{#2}
\providecommand{\BIBentrySTDinterwordspacing}{\spaceskip=0pt\relax}
\providecommand{\BIBentryALTinterwordstretchfactor}{4}
\providecommand{\BIBentryALTinterwordspacing}{\spaceskip=\fontdimen2\font plus
\BIBentryALTinterwordstretchfactor\fontdimen3\font minus \fontdimen4\font\relax}
\providecommand{\BIBforeignlanguage}[2]{{%
\expandafter\ifx\csname l@#1\endcsname\relax
\typeout{** WARNING: IEEEtran.bst: No hyphenation pattern has been}%
\typeout{** loaded for the language `#1'. Using the pattern for}%
\typeout{** the default language instead.}%
\else
\language=\csname l@#1\endcsname
\fi
#2}}
\providecommand{\BIBdecl}{\relax}
\BIBdecl

\bibitem{manias2024trends}
D.~M. Manias, A.~M. Saber, M.~I. Radaideh, A.~T. Gaber, M.~Maniatakos, H.~Zeineldin, D.~Svetinovic, and E.~F. El-Saadany, ``Trends in smart grid cyber-physical security: Components, threats and solutions,'' \emph{IEEE Access}, 2024.

\bibitem{kundur1994power}
P.~Kundur, \emph{Power System Stability and Control}.\hskip 1em plus 0.5em minus 0.4em\relax New York, NY, USA: McGraw-Hill, 1994.

\bibitem{khalaf2018joint}
M.~Khalaf, A.~Youssef, and E.~El-Saadany, ``Joint detection and mitigation of false data injection attacks in agc systems,'' \emph{IEEE Transactions on Smart Grid}, vol.~10, no.~5, pp. 4985--4995, 2018.

\bibitem{ozay2016ml}
M.~Ozay, I.~Esnaola, F.~T.~Y. Vural, S.~R. Kulkarni, and H.~V. Poor, ``Machine learning methods for attack detection in the smart grid,'' \emph{IEEE Transactions on Neural Networks and Learning Systems}, vol.~27, no.~8, pp. 1773--1786, 2016.

\bibitem{xu2019explainable}
F.~Xu, H.~Uszkoreit, Y.~Du, W.~Fan, D.~Zhao, and J.~Zhu, ``Explainable ai: A brief survey on history, research areas, approaches and challenges,'' in \emph{Natural language processing and Chinese computing: 8th cCF international conference, NLPCC 2019, dunhuang, China, October 9--14, 2019, proceedings, part II 8}.\hskip 1em plus 0.5em minus 0.4em\relax Springer, 2019, pp. 563--574.

\bibitem{lundberg2017shap}
S.~M. Lundberg and S.-I. Lee, ``A unified approach to interpreting model predictions,'' \emph{Advances in Neural Information Processing Systems}, vol.~30, 2017.

\bibitem{doshi2021explainable}
K.~Doshi, M.~Johnson, and G.~Lee, ``Explainable {AI}: The new frontier in decision support for power systems,'' \emph{Electric Power Systems Research}, vol. 194, p. 107009, 2021.

\bibitem{shi2024review}
H.~Shi, L.~Fang, X.~Chen, C.~Gu, K.~Ma, X.~Zhang, Z.~Zhang, J.~Gu, and E.~G. Lim, ``Review of the opportunities and challenges to accelerate mass-scale application of smart grids with large-language models,'' \emph{IET Smart Grid}, vol.~7, no.~6, pp. 737--759, 2024.

\bibitem{Chen2024}
M.~Chen, F.~Yang, H.~Liu, B.~Han, T.~Zhang, W.~Wu, and H.~V. Poor, ``Large language model based multi-objective artificial intelligence control of smart grids,'' \emph{IET Smart Grid}, vol.~7, no.~5, pp. 496--508, 2024.

\bibitem{Liu2024}
M.~Liu, Y.~Liu, T.~Zhao, H.~Yang, L.~Ge, and B.~Mather, ``Large language model for power system applications: Opportunities, challenges, and future directions,'' \emph{Electric Power Systems Research}, vol. 235, p. 110896, 2024.

\bibitem{Zhang2023}
Z.~Zhang, H.~Gong, C.~Li, Z.~Wang, L.~Xing, F.~Wang, H.~Zhang, and M.~Zhou, ``Empowering smart grid: A comprehensive review of machine learning, deep learning, and large language model applications,'' \emph{arXiv preprint arXiv:2305.11202}, 2023.

\bibitem{elgerd1982electric}
O.~I. Elgerd, \emph{Electric Energy Systems Theory: An Introduction}, 2nd~ed.\hskip 1em plus 0.5em minus 0.4em\relax New York, NY, USA: McGraw-Hill, 1982.

\bibitem{saber2024unmasking}
A.~M. Saber, A.~Youssef, D.~Svetinovic, H.~Zeineldin, and E.~F. El-Saadany, ``Unmasking covert intrusions: Detection of fault-masking cyberattacks on differential protection systems,'' \emph{IEEE Transactions on Systems, Man, and Cybernetics: Systems}, 2024.

\bibitem{akiba2019optuna}
T.~Akiba, S.~Sano, T.~Yanase, T.~Ohta, and M.~Koyama, ``Optuna: A next-generation hyperparameter optimization framework,'' in \emph{Proceedings of the 25th ACM SIGKDD international conference on knowledge discovery \& data mining}, 2019, pp. 2623--2631.

\end{thebibliography}




\end{document}